\documentclass[11pt]{article} 
\usepackage{graphicx,amssymb}  
\begin{document}  
 
\vskip 30pt

\begin{center}  
{\Large{\bf Smallness of $\theta_{13}$  
and the size of the Solar Mass Splitting: Are they related?}}\\
\vspace*{1cm}  
\renewcommand{\thefootnote}{\fnsymbol{footnote}}  
{ {\sf Soumita Pramanick${}$\footnote{email: soumitapramanick5@gmail.com}},  
{\sf Amitava Raychaudhuri${}$\footnote{email: palitprof@gmail.com}}
} \\  
\vspace{10pt}  
{\small  {\em Department of Physics, University of Calcutta,  
92 Acharya Prafulla Chandra Road, Kolkata 700009, India}}
\normalsize

\end{center}

\begin{abstract} 

Compared to the other neutrino mixing angles $\theta_{13}$ is
small. The solar mass splitting is about two orders smaller than
the atmospheric splitting.  We show that both could arise from a
peturbation of a more symmetric structure. The perturbation also
affects the solar mixing angle and can make alternate mixing
patterns such as tribimaximal, bimaximal, or other variants
equally viable.  For real perturbations this can be accomplished
only for normal mass ordering with the lightest neutrino mass
less than 10$^{-2}$ eV. Both mass orderings can be accommodated
by going over to complex perturbations if the lightest neutrino
is heavier.  The CP-phase in the lepton sector, fixed by
$\theta_{13}$ and the lightest neutrino mass,  distinguishes
between different mixing models. \\

\vskip 5pt \noindent  
\texttt{PACS No:~ 14.60.Pq}  \\  
\texttt{Key Words:~~Neutrino mixing, $\theta_{13}$, Solar
splitting, Perturbation}
\end{abstract}  

\renewcommand{\thesection}{\Roman{section}} 
\setcounter{footnote}{0} 
\renewcommand{\thefootnote}{\arabic{footnote}} 
\noindent

\section{Introduction}

The recent measurement \cite{DayaBay,RENO} of a non-zero
$\theta_{13}$ which is small compared to the other neutrino
mixing angles has created a stir in the world of particle
physics. 

The Daya Bay collaboration after 127 days exposure has obtained
for $\theta_{13}$ \cite{DayaBay}
\begin{equation}
\sin^22\theta_{13} =
0.089 \pm 0.010 \; {\rm (stat)} \pm 0.005\; {\rm (syst)} \;\;{\rm
(Daya ~Bay)}
\end{equation}
and from the RENO experiment with 229 days data \cite{RENO} one has
\begin{equation}
\sin^22\theta_{13}
= 0.113 \pm 0.013 \; {\rm (stat)} \pm 0.019\; {\rm (syst)} \;\;
{\rm (RENO)}
\end{equation}
The Double Chooz \cite{DChooz}, MINOS \cite{minos}, and T2K \cite
{t2k} experiments have also determined $\sin^22\theta_{13}$, all
consistent with the above but with larger uncertainties.

Earlier there already was in place a strong
upper bound on this angle \cite{pdg}.
The measured value is close to this limit, leading to
$\theta_{13}$ 
getting referred to occasionally as `large'.  
In terms of the three known mixing angles $\theta_{12}$,
$\theta_{23}$,  $\theta_{13}$ and a phase $\delta$ the
Pontecorvo, Maki, Nakagawa, Sakata (PMNS) mixing matrix is usually
parametrized as,
\begin{equation}
U=
\pmatrix{c_{12}c_{13} & s_{12}c_{13} & s_{13}e^{-i \delta} \cr
-s_{12}c_{23}-c_{12}s_{23}s_{13}e^{i \delta} & 
c_{12}c_{23}-s_{12}s_{23}s_{13}e^{i \delta} & 
s_{23}c_{13}
\cr
s_{12}s_{23}-c_{12}c_{23}s_{13}e^{i \delta} & 
-c_{12}s_{23}-s_{12}c_{23}s_{13}e^{i \delta} & 
c_{23}c_{13}
}.
\label{PMNS}
\end{equation}
As it is now realized that in the lepton sector, as for the
quarks, all three mixing angles are non-zero the door has been
opened for CP-violation\footnote{CP-violation in the heavy
neutrino sector could be the origin of matter-antimatter
asymmetry through leptogenesis.}.  Many alternative strategies
are being considered to explore leptonic CP-violation as well as
mixing and the future prospects are rich.

The other face of the neutrino sector is the mass spectrum.
Indeed, from the several oscillation studies at accelerators and
reactors complementing the solar and atmospheric neutrino
measurements the mass splittings are now very well established
though the absolute mass remains an unknown.
From global fits the currently favoured values of the neutrino
mixing parameters are
\cite{Gonzalez, Valle}:
\begin{eqnarray}
\Delta m_{21}^2 &=& (7.50 ^{+0.18}_{-0.19}) \times 10^{-5} \, {\rm eV}^2, \;\;
\theta_{12} = (33.36 ^{+0.81}_{-0.78})^\circ, \nonumber \\
|\Delta m_{31}^2| &=& (2.473 ^{+0.070}_{-0.067}) \times 10^{-3}
\, {\rm eV}^2, \;\;
\theta_{23} = (40.0 ^{+2.1}_{-1.5} \oplus 50.4 \pm 0.13)^\circ \nonumber \\
\theta_{13} &=& (8.66 ^{+0.44}_{-0.46})^\circ, \;\; \delta =
(300^{+66}_{-138})^\circ \;\;.
\label{results}
\end{eqnarray}

Note that the atmospheric mixing angle, $\theta_{23}$, is no
longer consistent with maximal mixing ($\theta_{23} = \pi/4$) at
1$\sigma$. There are best fit values in both the first and second
octants; determining the $\theta_{23}$ octant is one of the
priorities of future experiments.  In this work to simplify the
discussion and minimize parameters we will nonetheless take
$\theta_{23} = \pi/4$. We comment on the effect of the small
departure from maximality on the results. In the global fit
$\theta_{12}$ is also large but not maximal while $\theta_{13}$
is the smallest of the three.

For the solar sector the splitting, $\Delta m_{21}^2$, is known
in magnitude and sign while for the atmospheric neutrinos only
the magnitude, $|\Delta m_{31}^2|$, has been determined, the sign
remaining unknown\footnote{In \cite{Gonzalez} for inverted
ordering a best-fit value of $\Delta m_{32}^2$ has been given.
It is consistent   to within 1$\sigma$ with the best-fit
value of $|\Delta m_{31}^2|$ we have cited from their normal
ordering fits.}. Thus two options are left open, the {\em normal}
and the {\em inverted} ordering of the mass spectrum depending
upon whether this undetermined sign is positive or negative. One
noteworthy feature here is that the solar splitting is about two
orders of magnitude smaller than the atmospheric splitting:
$R_{\rm mass} = |\Delta m_{21}^2/\Delta m_{31}^2| = (3.03 \pm
0.16) \times 10^{-2}$.

The non-zero value of $\theta_{13}$ close to its upper bound
(`large') and yet small compared to the other mixing angles has
attracted a great deal of attention from diverse angles. We list
a sampling of this body of literature. For
example, the role of $\mu - \tau$ symmetry \cite{mutau}, see-saw
models \cite{seesaw}, charged lepton contributions \cite{CL}, and
renormalization group effects \cite{RG}
are among the avenues explored. A perturbative approach has been
espoused in \cite{pert}. Other attempts have been based on
diverse discrete symmetries \cite{disc1, disc2}.

In this work we seek to address the following question: Is it
possible that at some level the small quantities, the ratio
$R_{\rm mass}$ and $\theta_{13}$, are vanishing\footnote{This may
arise from a symmetry such as $O(2)$ \cite{Heeck}.} and that a single
perturbation induces the observed non-zero values for both? The answer is
in the affirmative. To our knowledge, this result was pointed out
for the first time through a specific
example in \cite{br}. Here, we make an exhaustive
analysis and show that the existence (or not) of a viable
solution depends on two factors: the ordering of the neutrino
masses and the mass of the lightest neutrino, $m_0$. For normal
ordering, for a large choice of parameters the requirements can
be met\footnote{An earlier work relating
$\theta_{13}$ to the solar oscillation parameters which favoured
normal mass ordering can be found in
\cite{akhmedov}.}. The perturbation can be real or complex. In the latter
case, CP-violation is present. The inverted ordering is less
favoured if the perturbation is real. In this case one would have
to admit significant differences in the sizes of the matrix
elements of the perturbation to get satisfactory solutions.

Our paper is structured as follows. In the next section we set up
the framework for our discussion and list some of the commonly
considered neutrino mixing schemes, e.g., tribimaximal mixing.
In the following section we elaborate on the degenerate
perturbative mechanism which we will adopt. Next we discuss to
what extent the global fits of the mixing parameters constrain
the choice of the perturbation. Our main results are presented in
the following section where we show the allowed ranges of the
perturbation matrix for the two mass orderings and the
predictions for CP-violation. We then briefly indicate how the
perturbation can arise from a mass model. We end with the
conclusions and discussions.

\section{Neutrino mass and mixing scenarios}

We restrict ourselves to the case of three flavours of neutrinos.
We also work in a basis where the charged lepton mass matrix is
diagonal. In this basis the entire lepton mixing resides in the
neutrino mass matrix.

Our starting point will be the unperturbed Majorana neutrino mass
matrix, $M^0$, which is always symmetric. We choose a form such that
the solar splitting is absent; i.e., in the mass basis one has
\begin{equation}
M^0_{mass} = diag (m^{(0)}_1, m^{(0)}_1, m^{(0)}_3).
\label{m0}
\end{equation}
For a specific mass ordering, the lightest neutrino mass, $m_0$,
determines $m^{(0)}_1$ and $m^{(0)}_3$. It is useful to
define\footnote{We take $m^{(0)}_i ~(i = 1,3)$ to be real and
positive. This can be accomplished by a suitable choice  of the
Majorana phases.} $m^{\pm} = (m^{(0)}_3 \pm m^{(0)}_1)$. $m^-$ is
positive (negative) for normal (inverted) mass ordering.

In the flavour basis the mass matrix becomes:
\begin{equation}
M^0_{flavour}= {U^0}\pmatrix{m^{(0)}_1 && \cr &m^{(0)}_1&\cr &&
m^{(0)}_3}{U^0}^T \;,
\label{m0flav}
\end{equation}
where $U^0$ is the lowest order 
leptonic mixing matrix. The columns of $U^0$ are the unperturbed
flavour eigenstates. Neutrino mass models lead to predictions
for $U^0$ of which three often-discussed variants are the
tribimaximal (TBM), bimaximal (BM), and the `golden ratio' (GR)
forms. Each of these imply $\theta_{13} = 0$ and $\theta_{23} =
\pi/4$. They differ only in $\theta_{12}$.  We will consider
them in turn along with a further option where there is no solar
mixing to start with. 

Our goal is to check whether in each case a perturbation mass
matrix, $M'$ (also symmetric),  can be identified which will add
corrections to $M^0$ and $U^0$ leading to mass splittings and
mixing angles in agreement with observations, in particular that
the correct $\Delta m_{21}^2$ and $\theta_{13}$ are realized.

\subsection{General parametrization}

In general as long as $\theta_{13} = 0$ and the atmospheric
mixing is maximal ($\theta_{23} = \pi/4$) the leptonic mixing matrix
can be parametrized as\footnote{This form has appeared earlier in
the literature, e.g. \cite{Ahluwalia}.}:
\begin{equation}
U^0=
\pmatrix{b & a & 0 \cr
-a/\sqrt{2} & b/\sqrt{2} & \sqrt{1 \over 2} \cr
a/\sqrt{2} & -b/\sqrt{2} & \sqrt{1 \over 2}},
\label{n4}
\end{equation}
with 
\begin{equation}
a^2 + b^2 = 1. 
\label{ab}
\end{equation}
For the above $U^0$ the solar
mixing angle is given by $\tan\theta^0_{12} = a/b$. The
experimentally determined range of $\theta_{12}$ in Eq.
(\ref{results}) corresponds to $0.539 \leq a \leq 0.561$  at 1$\sigma$. 

\subsection{Tribimaximal mixing}

The preferred values of the mixing angles are reasonably close to 
a mixing matrix of tribimaximal form \cite{hps},
\begin{equation}
U^0=
\pmatrix{\sqrt{2 \over 3} & \sqrt{1 \over 3} & 0 \cr
-\sqrt{ 1 \over 6} & \sqrt{ 1 \over 3} & \sqrt{1 \over 2} \cr
\sqrt{ 1 \over 6} & -\sqrt{ 1 \over 3} & \sqrt{1 \over 2}},
\label{n1}
\end{equation}
which predicts the third mixing angle $\theta_{13}$ to be exactly
vanishing. 

\subsection{Bimaximal mixing}

For bimaximal mixing the matrix is \cite{bm},
\begin{equation}
U^0=
\pmatrix{\sqrt{1 \over 2} & \sqrt{1 \over 2} & 0 \cr
-{1 \over 2} &  \frac{1}{2} & \sqrt{1 \over 2} \cr
{1 \over 2} & -{1 \over 2} & \sqrt{1 \over 2}},
\label{n2}
\end{equation}
which also has a vanishing $\theta_{13}$.

\subsection{Golden ratio mixing} 

A third form of the mixing matrix also appearing in the literature involves
the golden ratio $\phi = (1 + \sqrt{5})/2$ \cite{gr},
\begin{equation}
U^0=
\pmatrix{\sqrt{\phi \over \sqrt{5}} & \sqrt{1 \over \sqrt{5}\phi} & 0 \cr
-{1 \over \sqrt{2}}\sqrt{\frac{1}{\sqrt{5}\phi}} &  {1 \over
\sqrt{2}}\sqrt{\frac{\phi}{\sqrt{5}}}  & \sqrt{1 \over 2} \cr
{1 \over \sqrt{2}}\sqrt{\frac{1}{\sqrt{5}\phi}} & -{1 \over
\sqrt{2}}\sqrt{\frac{\phi}{\sqrt{5}}} & \sqrt{1 \over 2}},
\label{n3}
\end{equation}
which too gives $\theta_{13} = 0$. 

\subsection{No solar mixing} 
Finally, we also examine the possibility that the unperturbed
mixing matrix has $a$ = 0. This would imply one degenerate state
decoupled and the other maximally mixed to the third
(nondegenerate) state. For this choice $\theta_{12}^0$ = 0.
Another case with one decoupled degenerate state is $b$ = 0 for
which $\theta_{12}^0 = \pi/2$. These cases  give identical
physics results.

\begin{table}[tb]
\begin{center}
\begin{tabular}{|c|c|c| c|c|c|c|c|}
\hline
Mixing & \multicolumn{2}{|c|}{Global fit 1$\sigma$}
&\multicolumn{2}{|c|}{Global fit 3$\sigma$}&TBM &BM & GR\\
\cline{2-5}
parameter & $a_{min}$ & $a_{max}$ &  $a_{min}$ & $a_{max}$ & & & \\ \hline
$a$ & 0.539 & 0.561 & 0.515 & 0.585 & 0.577 & 0.707  & 0.526 \\ \hline
\end{tabular}
\end{center}
\caption{The limits on the mixing parameter $a \equiv
\sin\theta^0_{12}$ as obtained from the global fit. The values
of $a$ for the TBM, BM, and GR forms are also shown. }
\label{t1}
\end{table}

In Table \ref{t1} we list the allowed range of $a$ from the
global fit and its values in the TBM, BM, and the GR models. As
noted, the unperturbed matrix, $M^0$, is such that the solar
splitting is absent and two eigenvalues are degenerate. Due to
this degeneracy the two corresponding eigenstates are non-unique.
The perturbation, $M'$, which splits the degeneracy determines
the actual eigenstates which will be rotated with respect to the
first two columns of $U^0$ -- Eq. (\ref{n4}) -- by an angle
$\zeta$ also determined by $M'$.  Therefore, on inclusion of the
perturbation we have a resultant solar mixing angle given by
$\theta_{12} = \theta^0_{12} + \zeta$.


\section{Perturbation Strategy}

We will work in the mass basis unless explicitly mentioned
otherwise. Our discussion will involve only first order
perturbative corrections. The perturbation $M'$ is a $(3 \times
3)$ symmetric matrix which could be real or complex. These two
cases will be treated sequentially. The former provides a good
starting point for the latter.

After removing an irrelevant constant
part the perturbation, $M'$,  can be written as:
\begin{equation}
M' = m^+ \pmatrix{ 0 & \gamma & \xi \cr
\gamma &  \alpha & \eta \cr
\xi & \eta & \beta } \;\; .
\label{mpert}
\end{equation}

\subsection{Real Perturbation} 

In this case all entries in the  matrix $M'$ are real. For
perturbation theory to be acceptable the dimensionless entities
$\alpha, \beta, \gamma, \xi, \eta$ should be small compared to
unity.  Taken together with the unperturbed $M^0$ -- Eq.
(\ref{m0}) -- at lowest order the perturbation will induce the
solar oscillation parameters through $\alpha$ and $\gamma$;
$\theta_{13}$ will be determined by $\xi$ and $\eta$; while
$\beta$ will result in a small correction to $m^{(0)}_3$. 

\subsection{Complex Perturbation} 

If $M'$ is complex symmetric then it is not
hermitian\footnote{$M^0$ is hermitian by construction.}. In such
an event one takes the hermitian combination $(M^0 + M')^\dagger
(M^0 + M')$ and considers $M^{0\dagger} M^0$ as the unperturbed
term and $(M^{0\dagger} M' + M'^\dagger M^0)$ as the lowest order
perturbation.  The unperturbed eigenvalues will now be
$(m^{(0)}_i)^2$ and the perturbation matrix
\begin{equation}
(M^{0\dagger} M' + M'^\dagger M^0) = 
m^+ \pmatrix{ 0 & 2 m^{(0)}_1 {\rm Re}(\gamma) & 
m^{+} {\rm Re}(\xi) - i ~m^{-} {\rm Im}(\xi) \cr
2 m^{(0)}_1 {\rm Re}(\gamma) &  2 m^{(0)}_1 {\rm Re}(\alpha) & 
m^{+} {\rm Re}(\eta) - i ~m^{-} {\rm Im}(\eta) \cr
m^{+} {\rm Re}(\xi) + i ~m^{-} {\rm Im}(\xi) & 
m^{+} {\rm Re}(\eta) + i ~m^{-} {\rm Im}(\eta)  & 
2 m^{(0)}_3 {\rm Re}(\beta) } \;\; .
\label{cpert}
\end{equation}
The imaginary parts of $\alpha$, $\beta$, and $\gamma$ do not
appear in Eq. (\ref{cpert}). However, they do contribute at
higher order {\em via} the $M'^\dagger M'$ term.

\section{Relating elements of $M'$ to the data}

We look for solutions which are consistent with the global
neutrino parameter fits up to 1$\sigma$. In particular, the solar
mass splitting and $\theta_{13}$ must both emerge from the
perturbation.  We  discuss these aspects now.

\subsection{The solar mixing angle}

To lowest order, the solar mass splitting is obtained {\em via}
the $(2 \times 2)$ submatrix of the perturbation, $M'$, in the
space of the first two generations. 
For real $M'$ in terms of $r = \gamma/\alpha$ from eq.
(\ref{mpert}) this submatrix is:
\begin{equation}
M'_{(2\times 2)}=
m^+ \alpha \pmatrix{ 0 & r \cr r &  1 } \; {\rm for ~Real} ~M' \;.\;
\label{submr}
\end{equation}
If $M'$ is complex then  $r = {\rm Re}(\gamma)/{\rm Re}(\alpha)$
and 
\begin{equation}
(M^{0\dagger} M' + M'^\dagger M^0)_{(2\times 2)} = 
2 m^+ m^{(0)}_1 {\rm Re}(\alpha) \pmatrix{ 0 & r \cr r &  1 } \; 
{\rm for ~Complex} ~M'\;\; .
\label{submc}
\end{equation}
If $r = 0$ then $M'$ will produce a mass splitting but will not
change the solar mixing. For $r$ non-zero  the eigenstates are
rotated from those in $U^0$ through an angle $\zeta$ given by
\begin{equation}
\zeta = \frac{1}{2} \tan^{-1}(2r) \;, 
\label{zeta}
\end{equation}
independent of the prefactor of the matrix. As noted, the
tribimaximal, bimaximal
and golden ratio mixing models do not satisfy the currently
measured value of $\theta_{12}$ within 1$\sigma$. Therefore for
these cases we choose $r \neq 0$ in such a manner that when
the mass degeneracy is removed the mixing angle is tweaked to
within the allowed range. In Table 2 we show the ranges of $r$
for each of the three models that result in  $\theta_{12}$ values
consistent with observations. It is noteworthy that $r$ is small
in every case  (but for the $\theta_{12}^0 = 0$ alternative).
Since it is a ratio of two elements of the perturbation matrix it
could, in principle, be ${\cal O}$(1). The smallness can
be traced to the fact that as $r \rightarrow  0$  the mass matrix
in the flavour basis exhibits a $Z_2 \times Z_2$
symmetry\footnote{This is often a
subgroup of a larger symmetry such as $A(4)$.} of the
unperturbed model generated by:
\begin{equation}
U_1 = 
1 - 2~\pmatrix{ a^2 & ab/\sqrt{2} & -ab/\sqrt{2}\cr 
ab/\sqrt{2} & b^2/2 & -b^2/2\cr
-ab/\sqrt{2} & -b^2/2 & b^2/2 } \;\; {\rm and} \;\; 
U_2 = 
1 - 2~\pmatrix{ b^2 & -ab/\sqrt{2} & ab/\sqrt{2}\cr 
-ab/\sqrt{2} & a^2/2 & -a^2/2\cr
ab/\sqrt{2} & -a^2/2 & a^2/2 } \;.
\label{z2z2}
\end{equation}

\begin{table}[tb]
\begin{center}
\begin{tabular}{|c|c|c|c|c|c|c|c|c|}
\hline
Parameter & \multicolumn{2}{|c|}{TBM}& \multicolumn{2}{|c|}{BM}&
\multicolumn{2}{|c|}{GR}&\multicolumn{2}{|c|}{$\theta_{12}^0 = 0$}\\ 
\cline{2-9}
& $r_{min}$ & $r_{max}$ &  $r_{min}$ & $r_{max}$ &  $r_{min}$ &
$r_{max}$&  $r_{min}$ &
$r_{max}$ \\ \hline
$r ~(\times 10^2)$ & -4.59  & -1.95 &  -23.1 & -19.9 &  1.54 
 & 4.18 & 108 &  125 \\ \hline
\end{tabular}
\end{center}
\caption{The range of the off-diagonal entry, $r =
\gamma/\alpha$, in the 2$\times$2 submatrix of the perturbation
(see Eqs. (\ref{submr}, \ref{submc})) for the TBM, BM, and GR
alternatives that produces a $\theta_{12}$ consistent with the
global fits at 1$\sigma$. The $\theta_{12}^0 = 0$ alternative
is also noted. }
\label{t2}
\end{table}

Before closing this subsection it is worth noting that  
to lowest order in degenerate perturbation theory
the first two eigenstates are:
\begin{equation}
|\psi_1\rangle=
\cos \zeta \left[ \pmatrix{b \cr -a/\sqrt{2} \cr a/\sqrt{2} } - 
\bar{\xi} \pmatrix{0 \cr 1/\sqrt{2} \cr
1/\sqrt{2}}\right] - 
\sin \zeta \left[\pmatrix{a \cr b/\sqrt{2} \cr -b/\sqrt{2} } -
\bar{\eta} \pmatrix{0 \cr 1/\sqrt{2} \cr
1/\sqrt{2}} \right]\;,\;
\label{psi1}
\end{equation}
\begin{equation}
|\psi_2\rangle=
\sin \zeta \left[ \pmatrix{b \cr -a/\sqrt{2} \cr a/\sqrt{2} } - 
\bar{\xi} \pmatrix{0 \cr 1/\sqrt{2} \cr
1/\sqrt{2}}\right] + 
\cos \zeta \left[\pmatrix{a \cr b/\sqrt{2} \cr -b/\sqrt{2} } -
\bar{\eta} \pmatrix{0 \cr 1/\sqrt{2} \cr
1/\sqrt{2}} \right]\;,\;
\label{psi2}
\end{equation}
with $\zeta$ defined in Eq. (\ref{zeta}) and
\begin{equation}
\bar{\xi} = \left(\frac{m^+}{m^-}\right) \xi, \;\;
\bar{\eta} = \left(\frac{m^+}{m^-}\right) \eta \; {\rm for ~Real} ~M',
\label{xietar}
\end{equation}
and
\begin{equation}
\bar{\xi} = \left(\frac{m^+}{m^{-}}\right) {\rm Re}(\xi)  + 
i ~{\rm Im}(\xi),
\;\; \bar{\eta } = \left(\frac{m^+}{m^{-}}\right) {\rm Re}(\eta)  + 
i ~{\rm Im}(\eta) \; {\rm for ~Complex} ~M' \;.
\label{xietac}
\end{equation}

\subsection{The solar mass splitting}

The solar mass splitting is determined by the eigenvalues of the
submatrix in Eqs. (\ref{submr}) and (\ref{submc}). 

For real $M'$ the first order corrections to the degenerate eigenvalues are:
\begin{equation}
m^{(1)}_{2,1} = m^+ ~\frac{\alpha}{2} \left[ 1 \pm \sqrt{1 +
4r^2} \right] \;.
\end{equation}
Identifying the heavier eigenvalue with $m_2$, as required by the
solar data,  one has:
\begin{equation}
m^2_{2} - m^2_{1} = 2 m^+ m^{(0)}_1 \alpha \sqrt{1 + 4r^2} \;.
\label{solsplt}
\end{equation}
Up to small perturbative corrections $m^+
m^-$ gives the atmospheric mass splitting. Hence: 
\begin{equation}
R_{\rm mass} = |(m^2_{2} - m^2_{1})/(m^2_{3} - m^2_{1})| = 2
\frac{m^{(0)}_1}{|m^-|} ~\alpha \sqrt{1 + 4r^2} \;.
\label{solar}
\end{equation}
For complex $M'$ the corrections are to the squared masses and
one directly obtains  Eqs. (\ref{solsplt})  and (\ref{solar})
but for the
replacement $\alpha \rightarrow {\rm Re}(\alpha)$.

We will return to this equation when we discuss numerical
estimates of the element $\alpha$.

\subsection{Generating $\theta_{13} \neq 0$}

Using first order degenerate perturbation theory the corrected
wave-function $|\psi_3 \rangle$ is given by:
\begin{equation}
|\psi_3\rangle =
\pmatrix{0 \cr 1/\sqrt{2} \cr 1/\sqrt{2}} +
\bar{\xi}^* \pmatrix{b \cr -a/\sqrt{2} \cr a/\sqrt{2} } + 
\bar{\eta}^* \pmatrix{a \cr b/\sqrt{2} \cr -b/\sqrt{2} }  \;.
\label{psi3}
\end{equation}

To minimize the number of free parameters
we will restrict ourselves to only those perturbations which
leave the atmospheric mixing angle $\theta_{23}$ fixed at the
maximal value\footnote{We remark about deviations from maximal
mixing at the end.} of $\pi/4$. This gives the relationship:
\begin{equation}
\left(\frac{\bar{\xi}}{\bar{\eta}}\right)^*  = \frac{b}{a}  
\label{xieta}
\end{equation}
Since $a$ and $b$ are real Eq. (\ref{xieta})  implies that
$\bar\xi$ and $\bar\eta$ and hence $\xi$ and $\eta$ have the same
phase.  Comparing with Eq. (\ref{PMNS}) one then has:
\begin{equation}
\sin \theta_{13} ~e^{-i \delta} = \left[b~ \bar{\xi}^*  +
a ~\bar{\eta}^*  \right]  
= \frac{\bar{\xi}^*}{b} \;,
\label{th13}
\end{equation}
where we have used $(a^2 + b^2) = 1$.

For real $M'$ one has $\delta = 0$. Hence, from Eq. (\ref{xietar}) 
\begin{equation}
\xi = \left(\frac{m^-}{m^+}\right) b \sin \theta_{13}  \;.
\label{xir}
\end{equation}

In the next section, these formulae will be used to relate
$M'$ to the neutrino masses and mixings.

\section{Results}

We now have all the ingredients in place to determine the full
perturbation matrix and extract the consequences. Once the
neutrino mass ordering is chosen and the lightest neutrino mass,
$m_0$, specified, the unperturbed mass spectrum is fixed. The
matrix element  $\alpha$ is determined from the solar splitting
through Eq. (\ref{solar}).  The element $\beta$ makes a small
contribution (a few per cent) to the atmospheric neutrino
splitting and does not affect the physics at hand and so will not
be pursued any further in this section. $\gamma$ is fixed by the
ratio $r$ (see Table \ref{t2}).  Finally, $\xi$ and $\eta$ are
determined through Eqs.  (\ref{xieta}) - (\ref{xir}). The
question to be examined, for each of the popular mixing patterns
for both mass ordering options, is  for what range of $m_0$ are
these matrix elements of acceptable magnitude as a perturbation?

\subsection{Real perturbation}

First we consider  $M'$ real which
amounts to $\delta$  = 0 and  CP-conservation. In this case one can
determine the dependence of $\alpha$ on $m_0$ using Eq.
(\ref{solar}). Since $\xi$ and $\eta$ are proportional to each
other -- see Eq.  (\ref{xieta}) -- presenting any one of them is
adequate. Here we present $\xi$ (which is larger than (equal to)
$\eta$ for the TBM and GR (BM)  mixing models)  as a function of
$m_0$ as obtained from Eq.  (\ref{xir}).

\subsubsection{Normal mass ordering}

\begin{figure}[tb] 
\begin{center} 
{\includegraphics[scale=0.5,angle=0]{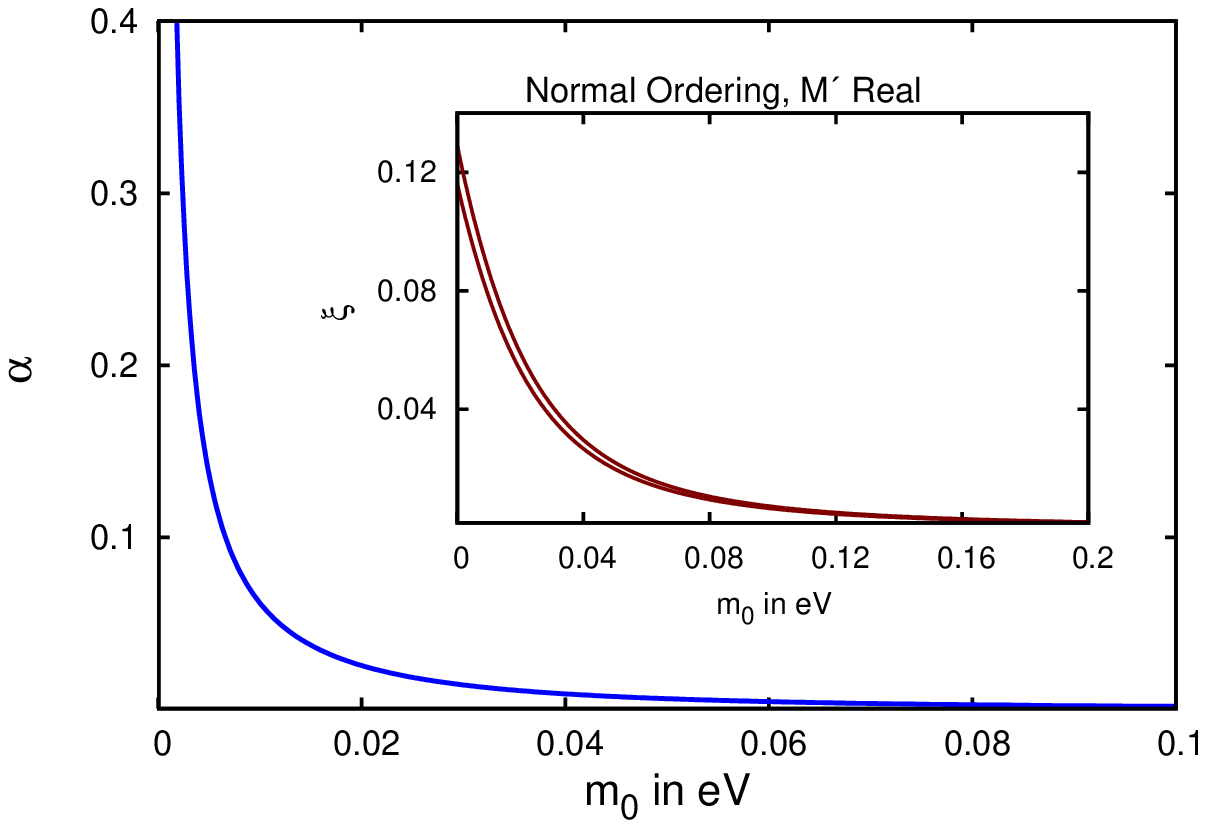}
\includegraphics[scale = 0.5, angle=0]{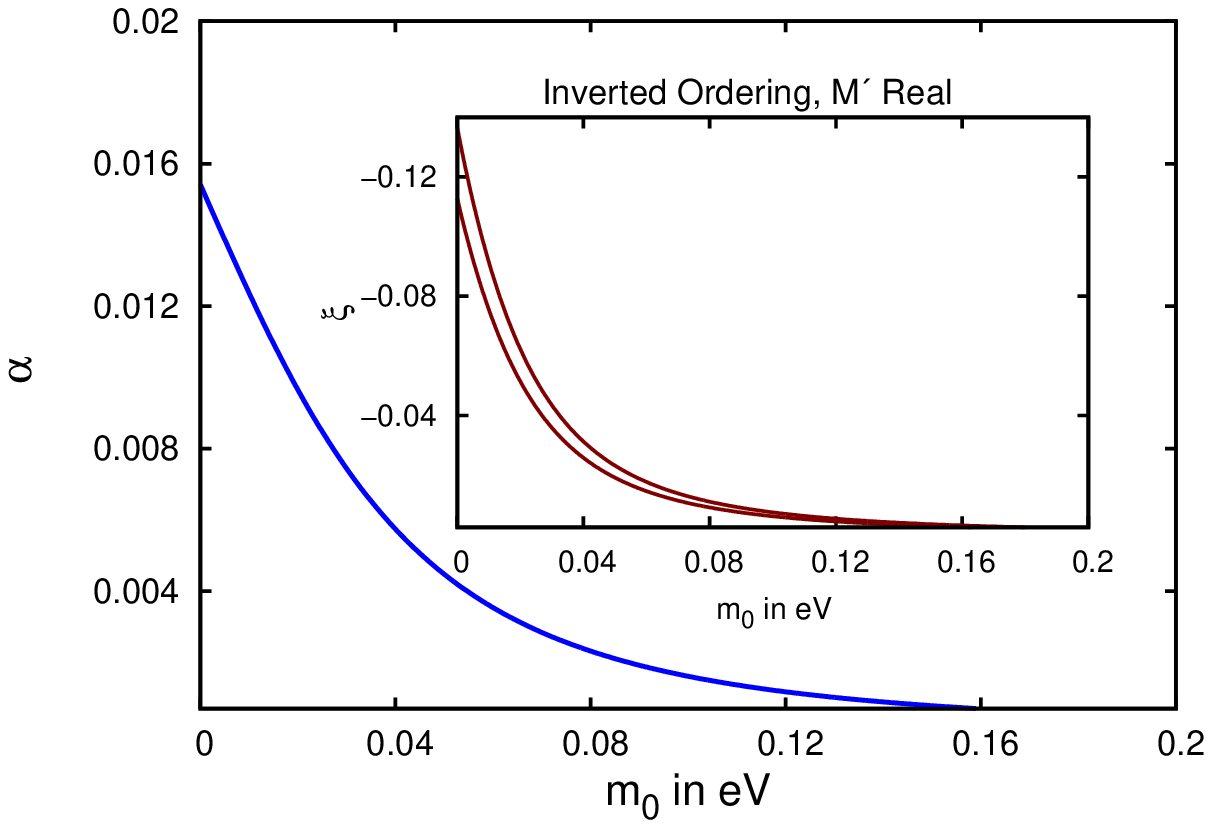}}
\caption{$\alpha$  and
$\xi$ (inset) as a function of the lightest neutrino mass
$m_0$ for real $M'$. The left (right) panel is for  normal
(inverted) mass ordering.  In the insets the region between
the two curves is allowed when $\theta_{13}$
is varied over its 1$\sigma$ range. The results for $\xi
~(\propto b)$ are for tribimaximal mixing ($b_{TBM} =
\sqrt{\frac{2}{3}} \sim 0.816$). The corresponding plots for
bimaximal ($b_{BM} = \frac{1}{\sqrt{2}} = 0.707$), `golden
ratio' mixing ($b_{GR} = \sqrt{\frac{\phi}{\sqrt{5}}} \sim
0.851$),  and the $\theta_{12}^0
= 0$ model ($b =1$) can be  obtained by scaling.}
\label{Real} 
\end{center} 
\end{figure} 

The results for normal ordering are in the left panel of Fig.
\ref{Real}. $\alpha$ is presented as a
function of the lightest neutrino mass $m_0$. We have shown the
case for $r$ = 0. We have verified that using the small values of
$r$ required to fit the solar mixing angle $\theta_{12}$ for the
popular models -- see Table \ref{t2} -- in Eq. (\ref{solar})
causes  no perceptible change\footnote{The corrections are ${\cal
O}(r^2)$.} in $\alpha$.  $r$ is larger for the $\theta_{12}^0
= 0$ model and this effectively reduces $\alpha$ by a factor of around
2.  As expected, $\alpha$ diverges as $m_0$ tends to zero.

In the inset we show $\xi$ as a function of $m_0$ for the
1$\sigma$  limits of
$\theta_{13}$.  In these plots  $b = \sqrt{\frac{2}{3}}$
corresponding to tribimaximal mixing. For the other commonly
considered alternatives -- bimaximal ($b = \frac{1}{\sqrt{2}}$)
and `golden ratio' ($b =\sqrt{\frac{\phi}{\sqrt{5}}}$) mixing --
the ordinate should be scaled appropriately. For the
$\theta_{12}^0 = 0$ model one must use $b = 1$.

\begin{figure}[bth] 
\begin{center} 
\includegraphics[scale=0.5,angle=0]{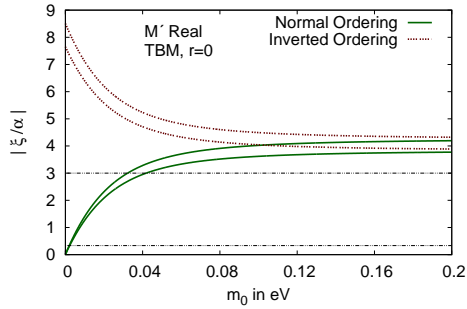}
\caption{The ratio $|\xi/\alpha|$ is plotted as function of the
lightest neutrino mass $m_0$ for both mass orderings when the
perturbation $M'$ is real. The area between the two curves of the
same type  is allowed  when $\theta_{13}$ is varied over its
1$\sigma$ range.  Also indicated are the values $1\over3$ and 3
for $|\xi/\alpha|$ -- black dot-dashed lines. }
\label{RealRatio} 
\end{center} 
\end{figure}

At this stage one can identify a favoured region of $m_0$ by
requiring that the elements of $M'$ -- such as $\xi$ and $\alpha$
-- should be of similar order.  For this
purpose, we plot in Fig. \ref{RealRatio} the ratio $|\xi/\alpha|$
as a function of $m_0$  (green solid curves).   For easy
identification we have shown where this ratio corresponds to the
values 3 and $\frac{1}{3}$ (dot-dashed black lines), two limits
separated by an order of magnitude.  Notice that for normal
ordering the ratio is within the above limits only if  $2.3
\times 10^{-3} ~{\rm eV} \leq m_0 \leq 3.7 \times 10^{-2} ~{\rm
eV} $.  If from other experiments a larger value of $m_0$ is
determined then that could be an indication that $M'$ must be
complex, as we discuss in the following section. We remind the
reader  that these curves are for tribimaximal mixing. For the
bimaximal (`Golden ratio') case the $\xi/\alpha$ curves will be
lowered (raised) by about 13.35 \% (4.28 \%). For the $\theta_{12}^0
= 0$ model $\alpha$ is reduced by a factor of about 2 while $\xi$
is enhanced by 25\%. As an upshot $\xi/\alpha$ is 2.5 times larger,
squeezing allowed $m_0$ to smaller values.

\subsubsection{Inverted mass ordering}

The results for inverted ordering appear in the right panel of Fig.
\ref{Real}. As before, $\alpha$ as a function of
$m_0$ is shown for  $r$ = 0 while $\xi$ for the 1$\sigma$ range
of $\theta_{13}$ is given in the inset. As for normal ordering,
inclusion in Eq. (\ref{solar}) of the small values of $r$ required
to achieve the best-fit $\theta_{12}$ in the TBM, BM, and GR
models causes essentially no change in $\alpha$. For the
$\theta_{12}^0 = 0$ model $\alpha$ is roughly halved. Once
again, we have used the TBM value  $b = \sqrt{\frac{2}{3}}$ for
the calculation of $\xi$. In this case $\xi$ turns out to be
negative. The two curves in the $\xi$ panel correspond to the
1$\sigma$ limits of $\theta_{13}$.

The noteworthy difference from normal ordering is that $\alpha$
is about an order of magnitude smaller than  $|\xi|$ for most of
the range of $m_0$. The brown dotted curves in Fig.
\ref{RealRatio}  depict the ratio $|\xi/\alpha|$ for
inverted ordering. It is seen that they lie outside the 
range of 1/3 to 3 for all  $m_0$ considered. Thus the inverted
ordering case would be a less favoured alternative for this
picture if the perturbation is real.

\subsection{Complex perturbation}

We now turn to the case of complex $M'$. If perturbation theory
is to be meaningful then we should expect the magnitudes of the
different dimensionless complex elements of $M'$ to be small
compared to unity. Barring fine tuning, they should also be of
roughly similar order.  Below, we take a conservative stand
and set: 
\begin{equation}
\alpha = \epsilon \exp(i\phi_\alpha), ~\gamma =
\epsilon \exp(i \phi_\gamma), ~\xi = \epsilon \exp(i\phi_\xi).
\end{equation}
The dimensionless quantity $\epsilon$ sets the scale of the
perturbation.  The phases $\phi_\alpha, \phi_\gamma$ and
$\phi_\xi$ are left arbitrary\footnote{The magnitude of $\eta$ is
determined through Eq. (\ref{xieta}). $\xi$ and $\eta$ have the
same phase.}.

It is seen from Eq. (\ref{xietac}) that
\begin{equation}
\tan \delta = \tan \phi_{\bar{\xi}} = \left(\frac{m^-}{m^+}\right) \tan{\phi_\xi}
\label{phases}
\end{equation}
where $\phi_{\bar{\xi}}$ is the phase of $\bar{\xi}$.

As we elaborate in the following, the phase freedom still leaves
room for some flexibility.  In particular, we will mostly focus
on those ranges of $m_0$ which are disfavoured for real $M'$ as
they do not satisfy the chosen criterion $3 \geq |\xi/\alpha|\geq
1/3$. We show that such $m_0$ are accommodated for complex $M'$.

\begin{figure}[bth] 
\begin{center} 
{\includegraphics[scale=0.5,angle=0]{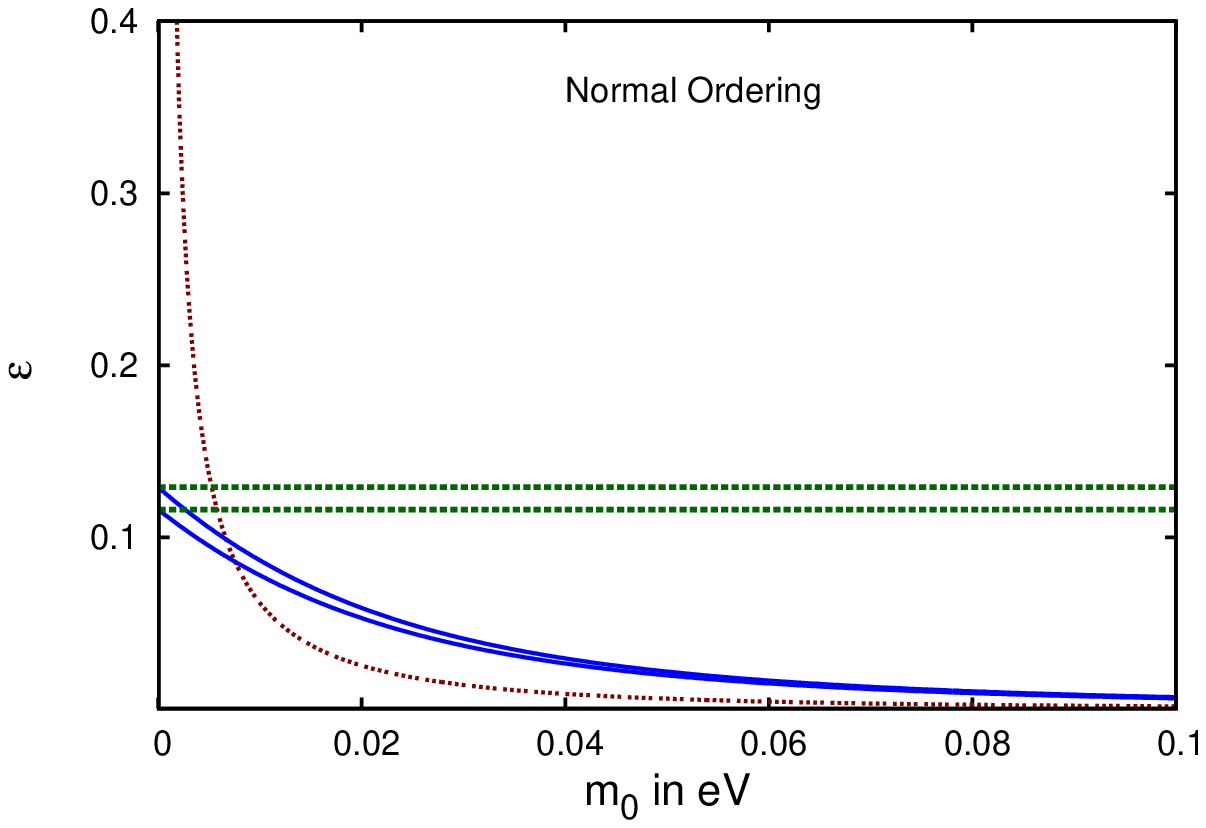}
\includegraphics[scale=0.5, angle=0]{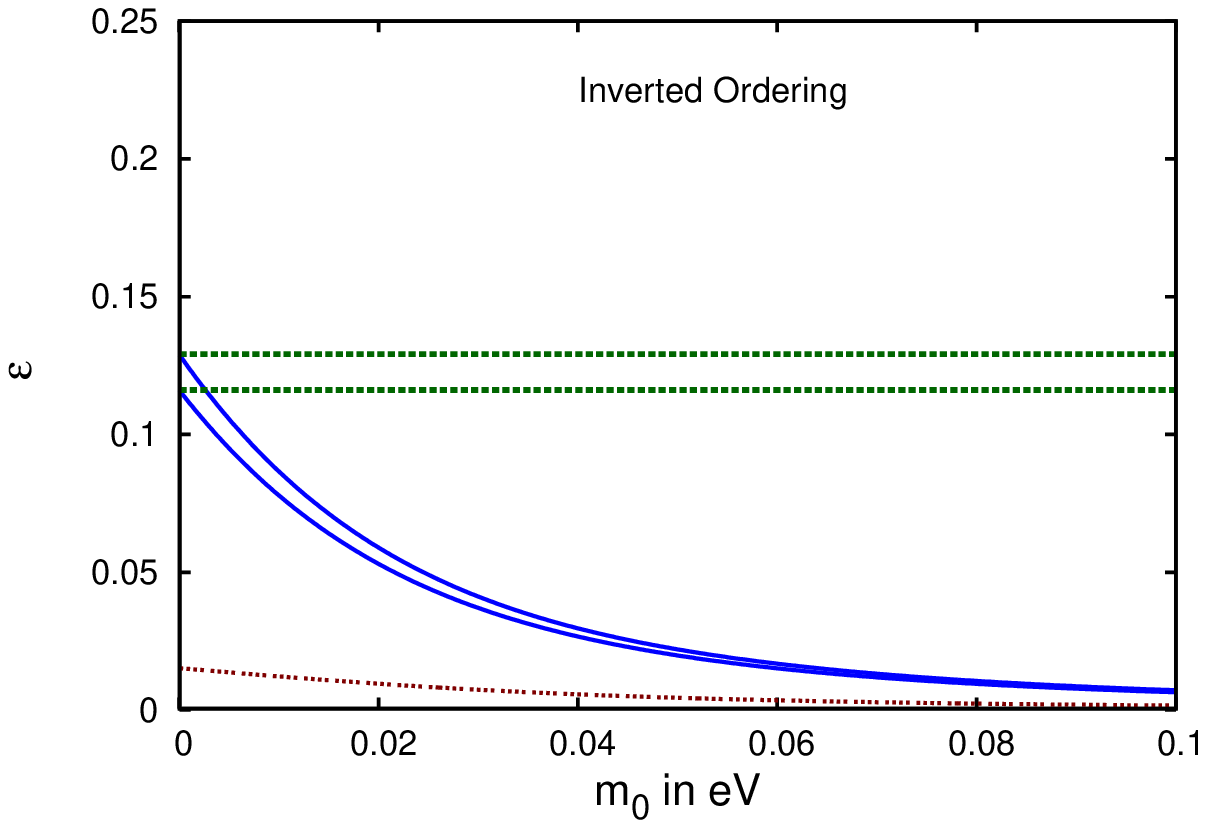}}
\caption{The limits on the scale of the perturbation,
$\epsilon$, for normal (left panel) and inverted (right panel)
mass orderings  as a function of the lightest neutrino mass
$m_0$. The upper (lower) limits from Eq. (\ref{xietal}) for
tribimaximal mixing are the green dashed (blue solid) curves.
The region between the curves of the same type  correspond to
$\theta_{13}$ values in the 1$\sigma$ range. The dotted maroon
curves are the lower limits from Eq.  (\ref{solarl}). Here $r$ =
0 has been taken.  }
\label{epslim} 
\end{center} 
\end{figure} 
The choice of $\epsilon$ is not entirely arbitrary. In particular,
Eq. (\ref{xietac}) implies: 
\begin{equation}
\left| \frac{m^+}{m^{-}} \right| ~\epsilon ~\geq ~|\bar{\xi}| ~\geq
~\epsilon \;\;.
\label{xietal}
\end{equation}
These limits are presented in the left (right) panel of Fig.
\ref{epslim} for the normal (inverted) mass ordering. The upper
and lower limits on $\epsilon$ are shown as the green dashed and
blue solid curves. The two curves of each type show how the limit
changes as $\theta_{13}$ is allowed to vary over its 1$\sigma$ 
range. Tribimaximal mixing has been assumed for these
plots.

In addition, from Eq. (\ref{solar}) one has
\begin{equation}
\epsilon \geq  \left|\frac{m^-}{2m^{(0)}_1 \sqrt{1 + 4r^2}}\right| 
R_{\rm mass}   \;.
\label{solarl}
\end{equation}
The lower limit from this equation is indicated by the dotted
maroon curves in the two panels of Fig. \ref{epslim}. This limit
is independent of both (a) the choice of $\theta_{13}$ and (b)
whether the mixing is of the tribimaximal, bimaximal, or Golden
ratio nature.  We have checked that the dependence on $r$ is
insignificant for the physics calculations.  It can be seen from
the left (right) panel of Fig. \ref{epslim} that for the normal
ordering (inverted ordering) for most values  of $m_0$
(for all values  $m_0$) the lower limit on $\epsilon$ from
$\bar{\xi}$ is more restrictive. Guided by these results, in the
following we choose $\epsilon$ = 0.1, 0.05, and
0.025.

\subsubsection{Normal mass ordering}

\begin{figure}[thb] 
\begin{center} 
{\includegraphics[scale=0.5,angle=0]{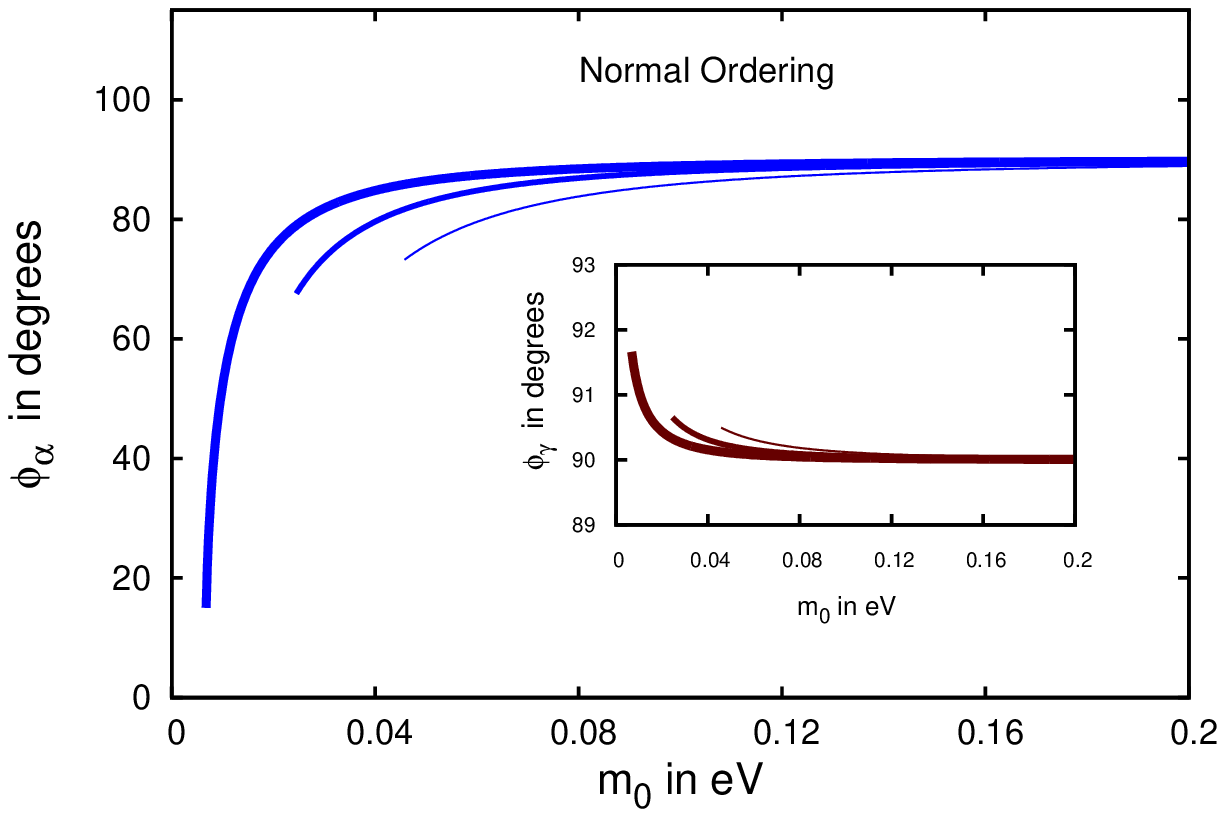}
\includegraphics[scale=0.5,angle=0]{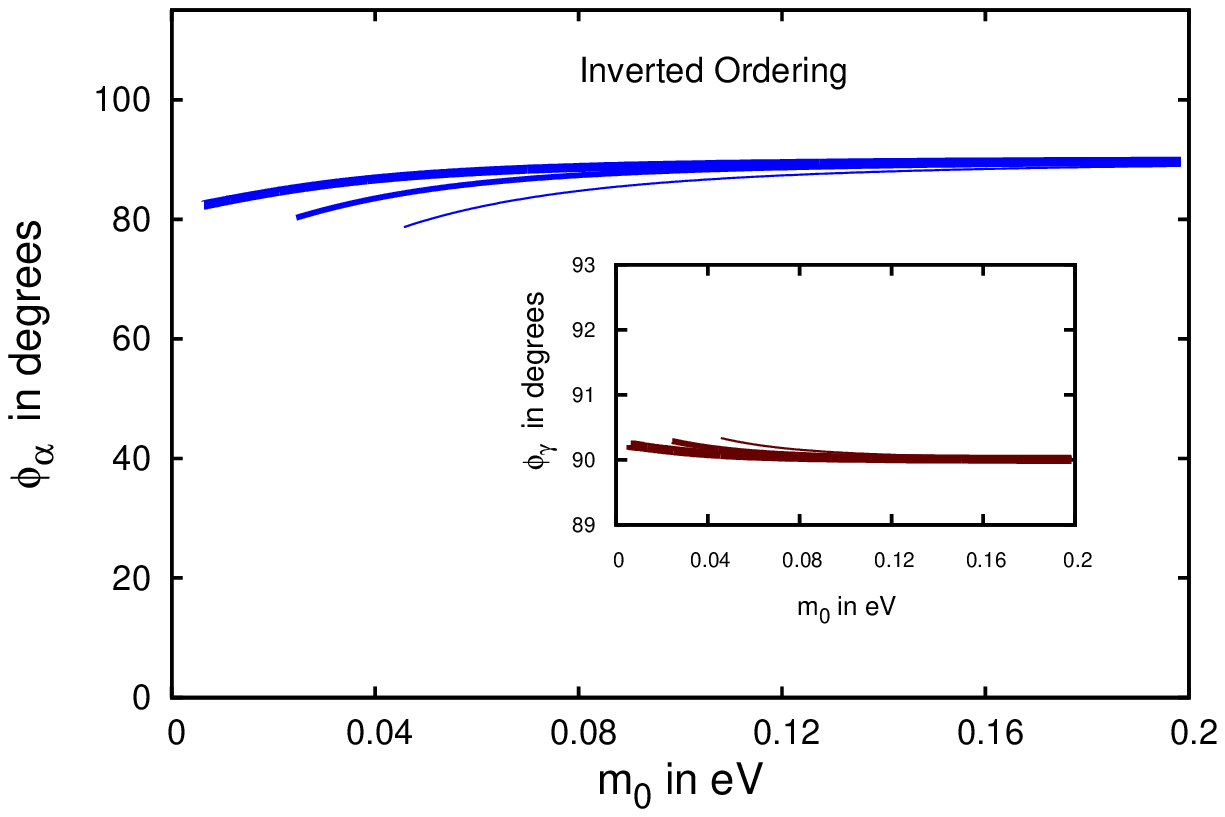}}
\caption{$\phi_\alpha$ ($\phi_\gamma$) for a complex $M'$ is
shown as a function of $m_0$ for normal mass ordering in the left
panel (left panel inset) for three values of $\epsilon$: in
decreasing order of line-thickness 0.1, 0.05 and 0.025.   In the
right panel  the same plots are displayed for inverted mass
ordering. }
\label{AlpGam} 
\end{center} 
\end{figure}

From Fig. \ref{RealRatio} it is seen that for real $M'$ and
normal mass ordering $|\xi/\alpha|$ is outside the chosen range for
$m_0 \geq 0.04$ eV. If $M'$ is complex, $\alpha$ in eq.
(\ref{solar}) is replaced by ${\rm Re}(\alpha)$.
Demanding that the solar splitting is correctly obtained 
fixes $\phi_\alpha$ when $\epsilon$ is chosen. The results are shown
in the left panel of Fig. \ref{AlpGam} for $\epsilon$ = 0.1, 0.05,
and 0.025. 

One can conclude from Fig. \ref{Real} that as $m_0$
increases ${\rm Re}(\alpha)$ approaches zero. This is reflected
in Fig. \ref{AlpGam} (left panel) where  $\phi_\alpha$ tends to
$\pi/2$ asymptotically for all choices of $\epsilon$. For a
particular $\epsilon$ the lightest neutrino mass 
$m_0$ has a lower limit set by Eq. (\ref{xietal}) where the
curves have been terminated. The corresponding $\phi_\alpha$ can be 
read off from Fig. \ref{epslim} -- $\cos \phi_\alpha$ is the
ratio of the value of the dot-dashed maroon curve to that of the
blue solid curve at this $m_0$. For these plots we have taken $r =
0$; the small corrections ${\cal O}(r^2)$ for the TBM, BM, and GR
models are insignificant. In the
$\theta_{12}^0 = 0$ model ${\rm Re}(\alpha)$ is reduced to about
half and so $\phi_\alpha$ tends closer to $\pi/2$.  One
should bear in mind that we have used the central value of
$R_{\rm mass}$ which has a $\pm 5$\% uncertainty.

As presented in Table \ref{t2}, in the TBM, BM, and GR models
the ratio $r = {\rm Re}(\gamma)/{\rm Re}(\alpha) = \cos
\phi_\gamma/\cos \phi_\alpha$ is tightly constrained from the
solar mixing angle $\theta_{12}$.  Thus $\phi_\gamma$ also tends
to $\pi/2$ as $m_0$ increases and since $r$ is small it does so
faster than $\phi_\alpha$. This can be seen from the inset in
Fig. \ref{AlpGam}.

\begin{figure}[thb]
\begin{center} 
{\includegraphics[scale=0.5, angle=0]{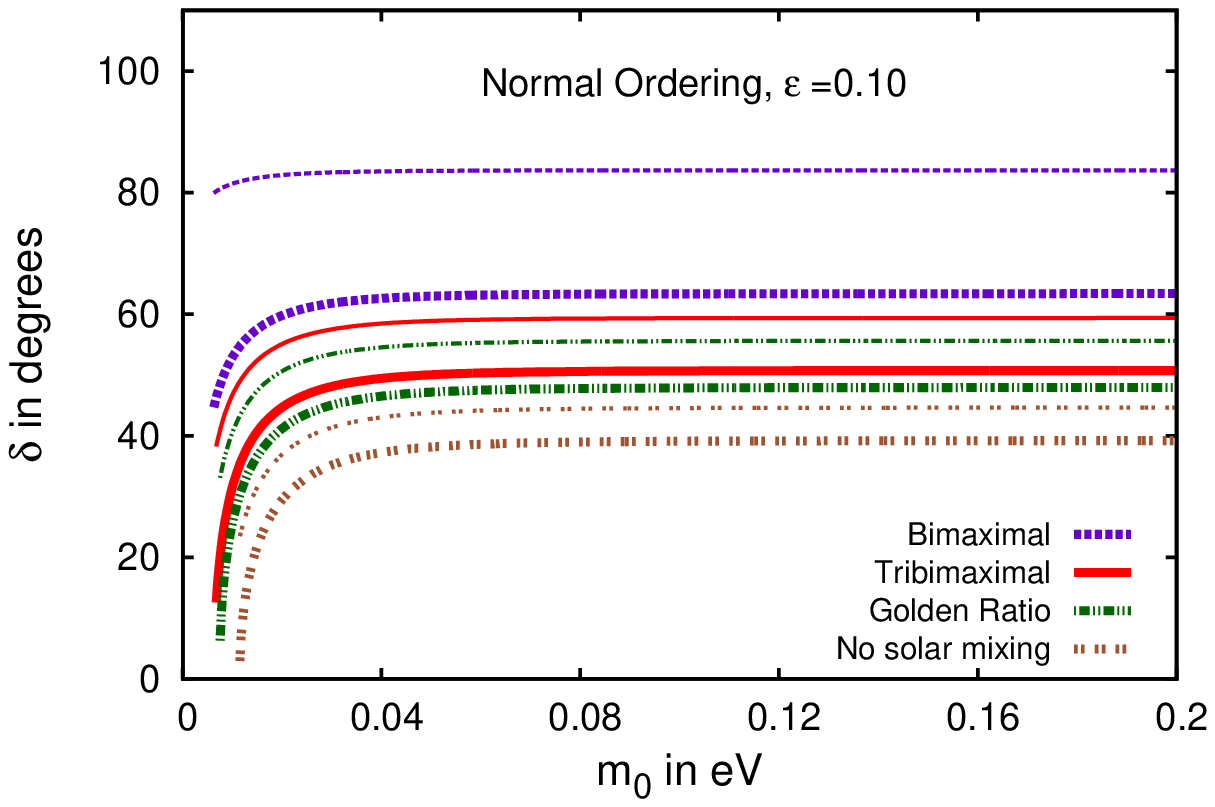}
\includegraphics[scale=0.5, angle=0]{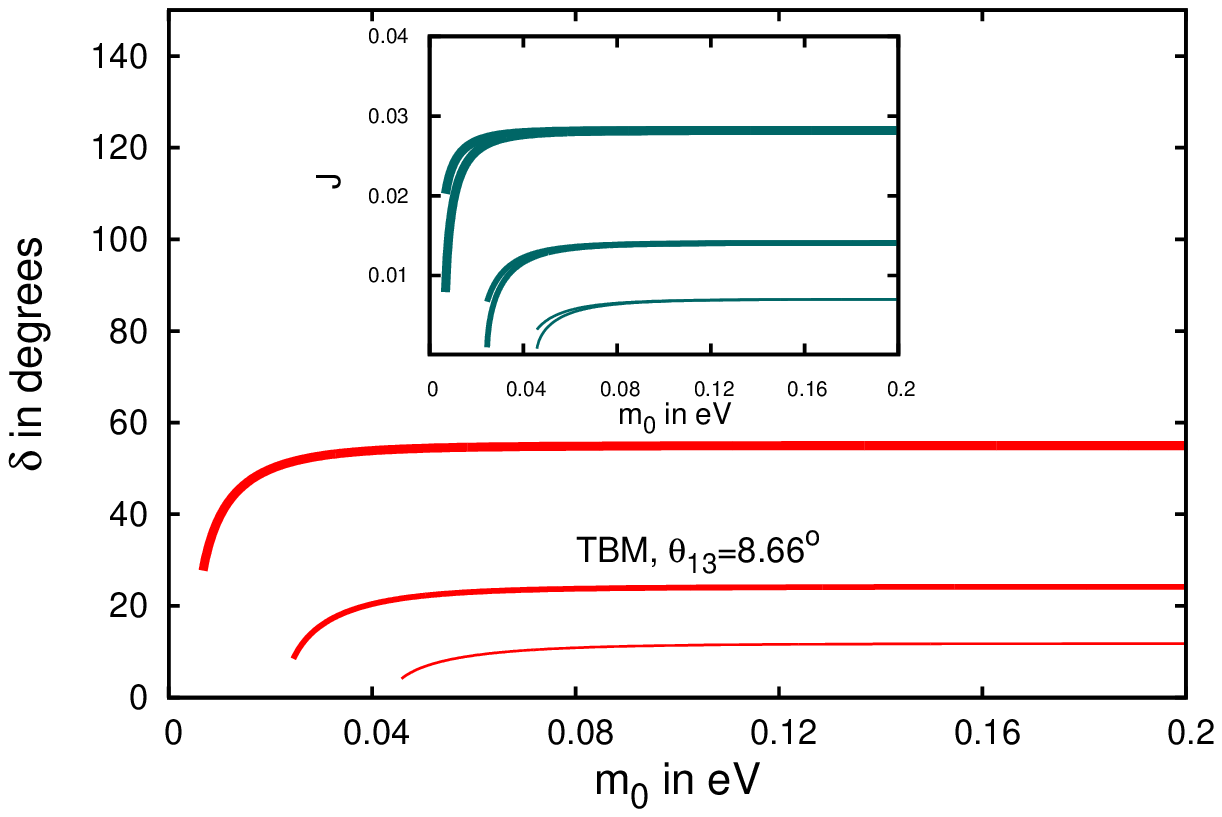}}
\caption{In the left panel $\delta$ for different models is
plotted for the 1$\sigma$ limiting values of $\theta_{13}$,
namely, $9.1^\circ$ (thick curves) and $8.2^\circ$ (thin curves).
$\epsilon$ has been taken to be 0.1.  The right panel is for the
TBM model.  Three values of $\epsilon$ are chosen -- in
decreasing order of thickness $\epsilon$ = 0.1, 0.05, 0.025 --
and $\theta_{13}$ is taken at the best-fit value. In the inset is
shown the Jarlskog parameter $J$ for the chosen $\epsilon$ and
the 1$\sigma$ limits of $\theta_{13}$. Both panels are for
normal mass ordering. For inverted ordering $\delta \rightarrow
(\pi -
\delta)$ and $J$  is unchanged.}
\label{DelJ} 
\end{center} 
\end{figure}

$\delta$ is not a free parameter in this model. Rather, 
picking a value for  $\theta_{13}$
amounts to fixing $|\bar{\xi}|$ from Eq. (\ref{th13}). Now, by
choice $|\xi| = \epsilon$, hence from Eq. (\ref{xietac}) one can
get $\phi_\xi$.  This in turn determines the phase of $\bar{\xi}$
which equals $\delta$. The results so obtained are presented in
the left panel of Fig. \ref{DelJ} for the TBM (red solid), BM
(violet dashed), and GR (green dot-dashed) models for $\epsilon =
0.1$.  The brown dotted curves are for $a = 0, b = 1$. For each
model the two curves correspond to the 1$\sigma$ upper and lower
limits of $\theta_{13}$.  It is worthwhile to point out that the
procedure for extracting $\delta$ using $|\bar{\xi}|$ leaves a
two-fold uncertainty $\delta \leftrightarrow \pi + \delta$.
Keeping this in mind we have shown $\delta$ in the first quadrant
in Fig. \ref{DelJ} even though the 1$\sigma$ range of the global
fit -- Eq. (\ref{results}) -- would prefer the partner $\pi +
\delta$ solution.

In the right panel of Fig. \ref{DelJ} we restrict to the case of
tribimaximal mixing and show the dependence of $\delta$ on the scale of
perturbation $\epsilon$. The conclusion that can be  drawn from
these panels is that $\delta$  is largely independent of the
lightest neutrino mass  and varies over a limited region
as $\theta_{13}$ covers its 1$\sigma$ range
or $\epsilon$ is varied. 

A reparametrization invariant measure of CP-violation is the
Jarlskog parameter, $J$ \cite{Jarlskog}. For 
arbitrary mixing it turns out to be 
\begin{equation}
J={\rm Im}[U_{e1}U_{\mu 2}U^*_{e2}U^*_{\mu 1}]=
\frac{1}{4b}\left[ (b^2 - a^2) \sin 2\zeta + 2 a b
\cos2\zeta \right] Im(\bar{\xi}) \;,
\label{Jarls}
\end{equation}
where in the last step only the lowest order perturbation effect
is retained.  In the inset of the
right panel of Fig \ref{DelJ} we show this CP-violation
measure as a function of $m_0$. Note that $J$ changes sign 
under $\delta \rightarrow \pi +
\delta$. We remark at this stage that for the $b = 0$ case $J$ = 0.

\subsubsection{Inverted mass ordering}

The analysis procedure  for inverted mass ordering is
essentially the same. In the right-panel of Fig. \ref{AlpGam} we
show $\phi_\alpha$ (with $\phi_\gamma$ in the inset) as a
function of $m_0$. The difference from  normal ordering arises
due to the appearance of $m_1^{(0)}$ in Eq. (\ref{solar})
which is larger in this case.
Hence,  $\phi_\alpha$ and
$\phi_\gamma$ remain closer to $\pi/2$ for all $m_0$.  The
determination of $\delta$ using Eqs. (\ref{th13}) and
(\ref{phases}) on the other hand involves only $m^\pm$ and not
$m_i^{(0)}$. Consequently, it can be seen from Eq. (\ref{phases})
that for any $m_0$ the CP-phase in 
the inverted and normal mass orderings are simply related  by $\delta
\leftrightarrow (\pi - \delta)$.
$J$ remains unchanged. So, $\delta$ and
$J$ for the inverted mass ordering can be read off from Fig.
\ref{DelJ}.

\section{Mass models}

The discussion thus far has not been tied to any specific model
for neutrino masses. We restrict ourselves to just a few remarks
here. The perturbation matrix in the flavour basis\footnote{As
noted, $\gamma$ is small compared to the scale of the
perturbation fixed by $\alpha, \xi$, and $\eta$. In this section,
we neglect $\gamma$.} corresponding to the general form of the
mixing matrix in Eq. (\ref{n4})  is:
\begin{equation}
M'_{flavour} = m^+ ~\pmatrix{a^2 \alpha & \chi & -
\sqrt{2} ab \alpha + \chi  \cr
\chi & (b^2 \alpha + \beta)/2  & (-b^2 \alpha + \beta)/2 \cr
- \sqrt{2} ab \alpha + \chi & (-b^2 \alpha + \beta)/2 &
(b^2 \alpha + \beta)/2}
\end{equation}
where use has been made of Eq. (\ref{xieta}) and we have set
$\chi = (ab \alpha +
b \xi + a \eta)/\sqrt{2}$. 

Attempting to relate the above matrix in its general form to the
popular mass models will take us beyond the scope of this paper.
Rather, we indicate here a limit when  it  can arise from a
Zee-type model \cite{zee}.  The required condition is:
\begin{equation}
\beta = \alpha (2 - 3b^2)
\label{alphabeta}
\end{equation} 
It is seen using Eq. (\ref{ab}) that for this choice the
diagonal elements of $M'_{flavour}$ become equal and can be
subsumed in the unperturbed matrix. The remaining terms can be
obtained from a Zee-type model\footnote{An alternate
derivation of $\theta_{13} \neq 0$ using the Zee model can be
found in \cite{HeMajee}.}. In these models $(M'_{flavour})_{ij}$ is
proportional to  $(M_i^2 - M_j^2)$, $M_i \;\; (i = 1,2,3)$
being the charged lepton masses. Since $m_\tau \gg m_\mu \gg
m_e$, without unnatural fine-tunings one would prefer $\chi$
to be much smaller than the other elements of the matrix. This
was already noted earlier \cite{br}; further details and
references can be found therein. An explicit $A(4)$-based model
which exhibits most of these features is given in \cite{disc2}.

For tribimaximal mixing, i.e., $a = \frac{1}{\sqrt{3}}$ and $b =
\sqrt{\frac{2}{3}}$, Eq. (\ref{alphabeta}) amounts to taking
$\beta = 0$. For
bimaximal mixing ($a = b = \frac{1}{\sqrt{2}}$) it is accomplished with
the choice  $\beta =
\alpha/2$. For the `golden ratio' mixing ($a = 
\sqrt{\frac{1}{\sqrt{5}\phi}}, b = 
\sqrt{\frac{\phi}{\sqrt{5}}}$) the choice
$\beta = \alpha (\phi - 2)/(2 \phi - 1)$ brings $M'_{flavour}$ to
the Zee form.

\section{Conclusion}

The neutrino mass spectrum and the mixing angles exhibit two
noteworthy features:  the mixing angle $\theta_{13}$ is small
compared to the the other two angles, namely, $\theta_{12}$ and
$\theta_{23}$, and the solar mass splitting is two orders of
magnitude smaller than the atmospheric splitting, $R_{\rm mass} =
|\Delta m_{21}^2/\Delta m_{31}^2| \simeq 10^{-2}$. We show that
both of these small quantities could be the result\footnote{An
attempt to generate the solar splitting and $\theta_{13}$ at low
energies starting from a partially degenerate mass spectrum and
$\theta_{13} = 0$ at a high scale through renormalisation group
effects in a supersymmetric model has been made in \cite{RG2}.}
of  a perturbation of a simpler partially degenerate neutrino
mass matrix ($m^{(0)}_1 = m^{(0)}_2$) along with a mixing matrix,
$U^0$,  which has $\theta_{13} = 0$.

The perturbation matrix can be chosen to be real only if the
neutrino mass ordering is normal and the lightest neutrino mass,
$m_0$, less than about 0.04 eV. In this case there will be no
CP-violation in the lepton sector.

For larger values of $m_0$ the pertubation $M'$ has to be
complex. We show that depending on the overall scale of the
perturbation, which we have indicated by $\epsilon$, the
CP-phase, $\delta$,  is calculable and could be near maximal
($\delta = \pi/2, 3\pi/2$) in some cases. CP-violation varies for
the different popular models -- e.g., tribimaximal, bimaximal,
`golden ratio', etc. It depends significantly on $\epsilon$ -- a
smaller perturbation resulting in a smaller $\delta$ -- but is
essentially independent of  $m_0$.  It also varies with
$\theta_{13}$ -- in the tribimaximal model the current 1$\sigma$
(3$\sigma$) range of $\theta_{13}$ translates to  about
10$^\circ$ (35$^\circ$) variation in $\delta$.

In this work we have taken the atmospheric mixing to be maximal
($\theta_{23} = \pi/4$). The current best-fit values are in the
two adjoining octants, both more than 1$\sigma$ away from
maximality. We have repeated the analysis using these two
best-fit values of $\theta_{23}$. We find that the CP-violation
effects are changed by less than 10\% in both
cases\footnote{Models have been proposed where the deviation of
$\theta_{23}$ from maximality is correlated with the value of
$\theta_{13}$ \cite{t23}.}.

As they stand, none of the popular mixing models are consistent
with the current value of $\theta_{12}$ at 1$\sigma$. We ensure
that for every model the perturbation takes care of this
shortcoming.  In passing, we also consider the possibility that
in the unperturbed case $\theta_{12} = 0 $ in addition to the
vanishing $\theta_{13}$. In such a scenario, both these angles
arise from the perturbation. In this case $\delta$ is the
smallest among all models.

Neutrino mass matrices which exhibit the features of the
unperturbed mass matrix are common in the literature. The
perturbative contribution can arise from a 
subdominant loop contribution from a Zee-type model. 

{\bf Acknowledgements:} SP acknowledges support from CSIR, India.
AR is partially funded by  the Department of Science and
Technology Grant No. SR/S2/JCB-14/2009.




\begin{thebibliography}{100} 



\bibitem{DayaBay}
  D.~A.~Dwyer [Daya Bay Collaboration],
  Nucl.\ Phys.\ Proc.\ Suppl.\  {\bf 235-236}, 30 (2013)
  [arXiv:1303.3863 [hep-ex]].
  F.~P.~An {\it et al.}  [DAYA-BAY Collaboration],
  Phys.\ Rev.\ Lett.\  {\bf 108}, 171803 (2012)
  [arXiv:1203.1669 [hep-ex]].

\bibitem{RENO} 
  J.~K.~Ahn {\it et al.}  [RENO Collaboration],
  Phys.\ Rev.\ Lett.\  {\bf 108}, 191802 (2012)
  [arXiv:1204.0626 [hep-ex]].


\bibitem{DChooz}
  Y.~Abe {\it et al.}  [DOUBLE-CHOOZ Collaboration],
  Phys.\ Rev.\ Lett.\  {\bf 108}, 131801 (2012)
  [arXiv:1112.6353 [hep-ex]]; 
  Phys.\ Lett.\ B {\bf 723}, 66 (2013)
  [arXiv:1301.2948 [hep-ex]].


\bibitem{minos}
  P.~Adamson {\it et al.}  [MINOS Collaboration],
  Phys.\ Rev.\ Lett.\ 
  [arXiv:1301.4581 [hep-ex]].


\bibitem{t2k}
  K.~Abe {\it et al.}  [T2K Collaboration],
  arXiv:1304.0841 [hep-ex].

\bibitem{pdg} The Review of Particle Physics,
K. Nakamura et al. (Particle Data Group), J. Phys.\ G {\bf 37}, 075021 (2010).

\bibitem{Gonzalez} 
  M.~C.~Gonzalez-Garcia, M.~Maltoni, J.~Salvado and T.~Schwetz,
  JHEP {\bf 1212}, 123 (2012)
  [arXiv:1209.3023v3 [hep-ph]].

\bibitem{Valle}
  D.~V.~Forero, M.~Tortola and J.~W.~F.~Valle,
  Phys.\ Rev.\ D {\bf 86}, 073012 (2012)
  [arXiv:1205.4018 [hep-ph]].
 


\bibitem{mutau}  
  J.~Liao, D.~Marfatia and K.~Whisnant,
  Phys.\ Rev.\ D {\bf 87}, 013003 (2013)
  [arXiv:1205.6860 [hep-ph]];
  S.~Gupta, A.~S.~Joshipura and K.~M.~Patel,
  arXiv:1301.7130 [hep-ph];
  A.~Damanik,
  arXiv:1305.6900 [hep-ph].




\bibitem{seesaw}  
  B.~Grinstein and M.~Trott,
  JHEP {\bf 1209}, 005 (2012)
  [arXiv:1203.4410 [hep-ph]];
  D.~Borah and M.~K.~Das,
  Nuclear Physics B 870, {\bf 461} (2013)
  [arXiv:1210.5074 [hep-ph]].





\bibitem{CL} 
  D.~Meloni, F.~Plentinger and W.~Winter,
Phys.\ Lett.\ B {\bf 699}, 354 (2011)
[arXiv:1012.1618 [hep-ph]];
  D.~Marzocca, S.~T.~Petcov, A.~Romanino and M.~Spinrath,
JHEP {\bf 1111}, 009 (2011)
[arXiv:1108.0614 [hep-ph]];
  C.~Duarah, A.~Das and N.~N.~Singh,
  Phys.\ Lett.\ B {\bf 718}, 147 (2012)
  [arXiv:1207.5225 [hep-ph]];
  S.~Gollu, K.~N.~Deepthi and R.~Mohanta,
  arXiv:1303.3393 [hep-ph].


\bibitem{RG}
  S.~Boudjemaa and S.~F.~King,
Phys.\ Rev.\ D {\bf 79}, 033001 (2009)
[arXiv:0808.2782 [hep-ph]];
  S.~Goswami, S.~T.~Petcov, S.~Ray and W.~Rodejohann,
Phys.\ Rev.\ D {\bf 80}, 053013 (2009)
[arXiv:0907.2869 [hep-ph]].

\bibitem{pert} 
  D.~Aristizabal Sierra, I.~de Medeiros Varzielas and E.~Houet,
  Phys.\ Rev.\ D {\bf 87}, 093009 (2013)
  [arXiv:1302.6499 [hep-ph]];
  R.~Dutta, U.~Ch, A.~K.~Giri and N.~Sahu,
  arXiv:1303.3357 [hep-ph];
  L.~J.~Hall and G.~G.~Ross,
  arXiv:1303.6962 [hep-ph];
  B.~Adhikary, A.~Ghosal and P.~Roy,
  arXiv:1210.5328 [hep-ph].


\bibitem{disc1}
  G.~Altarelli and F.~Feruglio,
Nucl.\ Phys.\ B {\bf 741}, 215 (2006)
[hep-ph/0512103];
  E.~Ma and D.~Wegman,
Phys.\ Rev.\ Lett.\  {\bf 107}, 061803 (2011)
[arXiv:1106.4269 [hep-ph]];
  S.~Gupta, A.~S.~Joshipura and K.~M.~Patel,
Phys.\ Rev.\ D {\bf 85}, 031903 (2012)
[arXiv:1112.6113 [hep-ph]];
  S.~Dev, R.~R.~Gautam and L.~Singh,
Phys.\ Lett.\ B {\bf 708}, 284 (2012)
[arXiv:1201.3755 [hep-ph]].
  G.~C.~Branco, R.~Gonzalez Felipe, F.~R.~Joaquim and H.~Serodio,
  Phys.\ Rev.\ D {\bf 86}, 076008 (2012)
  [arXiv:1203.2646 [hep-ph]];
  E.~Ma,
Phys.\ Lett.\ B {\bf 660}, 505 (2008)
[arXiv:0709.0507 [hep-ph]].
  F.~Plentinger, G.~Seidl and W.~Winter,
JHEP {\bf 0804}, 077 (2008)
[arXiv:0802.1718 [hep-ph]];
  N.~Haba, R.~Takahashi, M.~Tanimoto and K.~Yoshioka,
Phys.\ Rev.\ D {\bf 78}, 113002 (2008)
[arXiv:0804.4055 [hep-ph]];
  S.~-F.~Ge, D.~A.~Dicus and W.~W.~Repko,
Phys.\ Rev.\ Lett.\  {\bf 108}, 041801 (2012)
[arXiv:1108.0964 [hep-ph]];
  T.~Araki and Y.~F.~Li,
Phys.\ Rev.\ D {\bf 85}, 065016 (2012)
[arXiv:1112.5819 [hep-ph]];
  Z.~-z.~Xing,
  Chin.\ Phys.\ C {\bf 36}, 281 (2012)
  [arXiv:1203.1672 [hep-ph]];
Phys.\ Lett.\ B {\bf 696}, 232 (2011)
[arXiv:1011.2954 [hep-ph]];
  P.~S.~Bhupal Dev, B.~Dutta, R.~N.~Mohapatra and M.~Severson,
  Phys.\ Rev.\ D {\bf 86}, 035002 (2012).



\bibitem{disc2} 
  B.~Adhikary, B.~Brahmachari, A.~Ghosal, E.~Ma and M.~K.~Parida,
Phys.\ Lett.\ B {\bf 638}, 345 (2006)
[hep-ph/0603059].


\bibitem{Heeck} 
  J.~Heeck and W.~Rodejohann,
  JHEP {\bf 1202}, 094 (2012)



\bibitem{br} 
  B.~Brahmachari and A.~Raychaudhuri,
  Phys.\ Rev.\ D {\bf 86}, 051302 (2012)
  [arXiv:1204.5619 [hep-ph]].

\bibitem{akhmedov}
Evgeny K. Akhmedov, G.C. Branco, M.N. Rebelo, Phys.\ Rev.\ Lett.\ {\bf 84}, 3535 (2000)
[hep-ph/9912205].


\bibitem{Ahluwalia} 
  I.~Stancu and D.~V.~Ahluwalia,
  Phys.\ Lett.\ B {\bf 460}, 431 (1999).


\bibitem{hps}
See, for example, 
P.F. Harrison, D.H. Perkins and W.G. Scott, Phys. Lett.\ B {\bf
530}, 167 (2002); Z. -z. Xing, Phys.  Lett.\ B {\bf 533}, 85
(2002); X. He and A. Zee, Phys. Lett.\ B {\bf 560}, 87 (2003).

\bibitem{bm} 
See, for example, 
  F.~Vissani,
  hep-ph/9708483;
  V.~D.~Barger, S.~Pakvasa, T.~J.~Weiler and K.~Whisnant,
  Phys.\ Lett.\ B {\bf 437}, 107 (1998)
  [hep-ph/9806387];
  R.~N.~Mohapatra and S.~Nussinov,
  Phys.\ Rev.\ D {\bf 60}, 013002 (1999)
  [hep-ph/9809415].


\bibitem{gr} 
See, for example, 
  A.~Datta, F.~-S.~Ling and P.~Ramond,
  Nucl.\ Phys.\ B {\bf 671}, 383 (2003)
  [hep-ph/0306002];
  Y.~Kajiyama, M.~Raidal and A.~Strumia,
  Phys.\ Rev.\ D {\bf 76}, 117301 (2007)
  [arXiv:0705.4559 [hep-ph]];
  G.~-J.~Ding, L.~L.~Everett and A.~J.~Stuart,
  Nucl.\ Phys.\ B {\bf 857}, 219 (2012)
  [arXiv:1110.1688 [hep-ph]].


\bibitem{Jarlskog}
  C.~Jarlskog,
  Phys.\ Rev.\ Lett.\  {\bf 55}, 1039 (1985);
  O.~W.~Greenberg,
  Phys.\ Rev.\ D {\bf 32}, 1841 (1985).



\bibitem{zee} 
A. Zee, Phys. Lett.\ B {\bf 93}, 389 (1980), Erratum-ibid. {\bf 95},
461 (1980). 

\bibitem{HeMajee} 
  X.~-G.~He and S.~K.~Majee,
JHEP {\bf 1203}, 023 (2012)
[arXiv:1111.2293 [hep-ph]].

\bibitem{RG2} 
  M.~Borah, B.~Sharma and M.~K.~Das,
  arXiv:1304.0164 [hep-ph].


\bibitem{t23}
  T.~Araki,
  arXiv:1305.0248 [hep-ph];
  M.~-C.~Chen, J.~Huang, K.~T.~Mahanthappa and A.~M.~Wijangco,
  arXiv:1307.7711 [hep-ph].





\end{thebibliography}
\end{document}